\documentclass[aps,prl,nofootinbib,amsmath,twocolumn,preprintnumbers,superscriptaddress]{revtex4-1}
\usepackage{amssymb,esvect,amsmath,graphicx,latexsym,amsthm,slashed,eso-pic,hyperref}
\usepackage{float}
\usepackage[hang,flushmargin]{footmisc}
\usepackage{placeins}
\usepackage{bbold}
\usepackage{multirow}
\usepackage{pifont}
\usepackage{slashed}
\usepackage{url}
\usepackage{cancel}
\usepackage[normalem]{ulem}

\begin{document}

\widetext

\title{Doped Semiconductor Devices for sub-MeV Dark Matter Detection}

\author{Peizhi Du}
\affiliation{New High Energy Theory Center, Rutgers University, Piscataway, NJ 08854, USA}

\author{ Daniel Ega\~na-Ugrinovic}
\affiliation{Perimeter Institute for Theoretical Physics, Waterloo, ON N2L 2Y5}

\author{Rouven Essig}
\affiliation{C.N. Yang Institute for Theoretical Physics, Stony Brook University, Stony Brook, NY, 11794, USA}

\author{Mukul Sholapurkar}
\affiliation{Department of Physics, University of California, San Diego, CA 92093, USA}

\preprint{YITP-SB-2022-41}
 \begin{abstract} 
 Dopant atoms in semiconductors can be ionized with $\sim10$ meV energy depositions, allowing for the design of low-threshold detectors. 
 We propose using doped semiconductor targets to search for sub-MeV dark matter scattering or sub-eV dark matter absorption on electrons. 
 Currently unconstrained cross sections could be tested with a 1 g-day exposure in a doped detector with backgrounds at the level of existing pure semiconductor detectors, but improvements would be needed to probe the freeze-in target. We discuss the corresponding technological requirements and lay out a possible detector design. 
 
 \end{abstract}
 
  \maketitle

\section{Introduction}

A variety of strategies have been explored and proposed to directly detect dark matter (DM) via the interactions that it may have with the Standard Model (SM) particles. One possibility is to design detectors that look for excitations of a material as the DM scatters off or is absorbed in it. The sensitivity of such experiments at low DM masses is fundamentally limited by the minimum energy required to create an excitation in the material. Currently, several leading constraints on DM with masses below a GeV are set by detectors that look for DM-electron interactions using semiconductor targets, where thresholds are set by the ionization energy of the valence-band electrons (the bandgap), which is typically of order $1$~eV~\cite{Essig:2011nj,Tiffenberg:2017aac,Crisler:2018gci,Agnese:2018col,Abramoff:2019dfb,Aguilar-Arevalo:2019wdi,SENSEI:2020dpa,Arnaud:2020svb,Amaral:2020ryn,DAMIC-M:2022aks}. This is sensitive to DM scattering or absorption for DM masses exceeding $\sim 1$~MeV or $\sim  1$~eV, respectively. 

\begin{figure}[t!]
\includegraphics[width=0.5\textwidth]{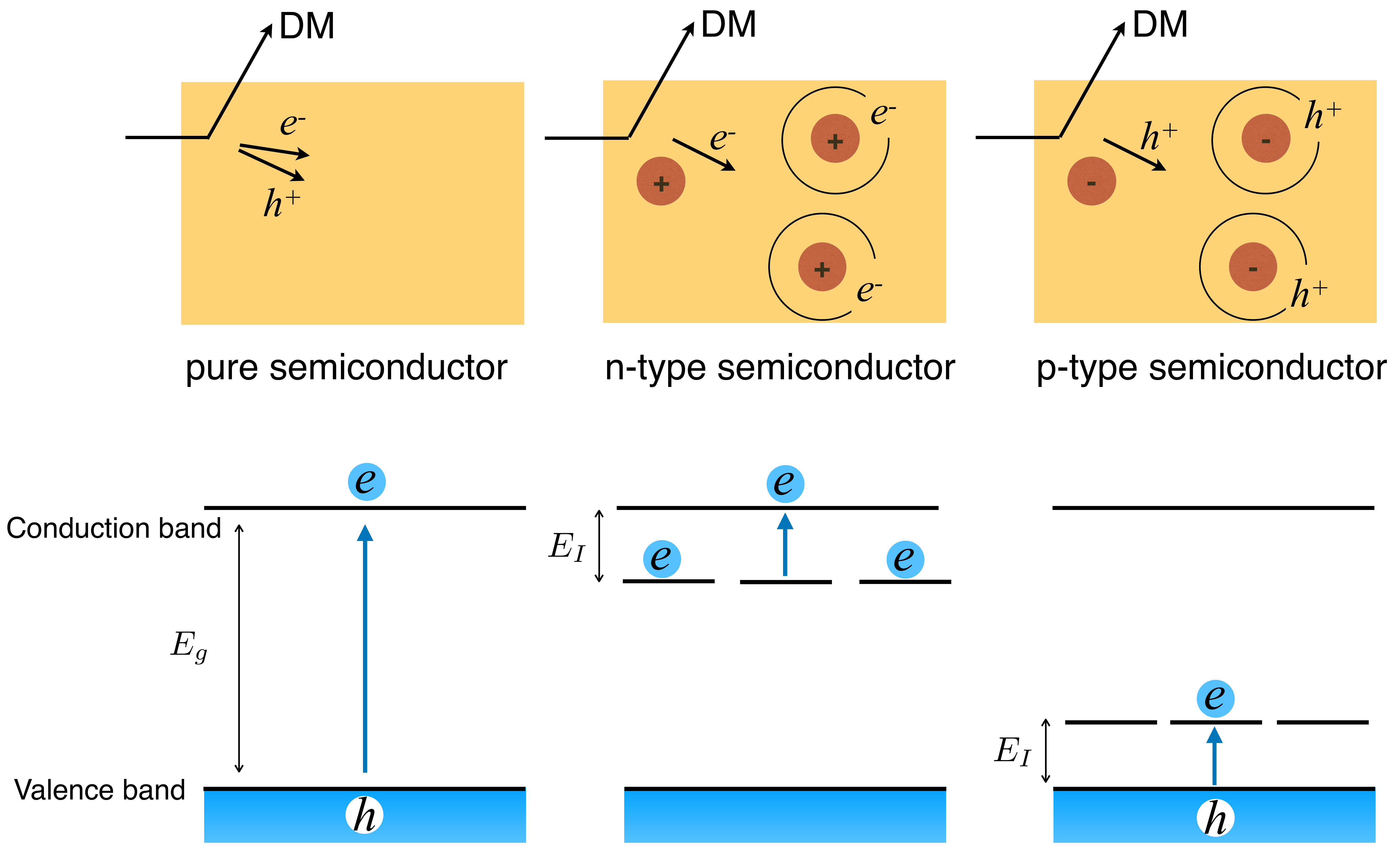}

\caption{Pure and doped semiconductor ionization signals: electron-hole pair for pure semiconductors (left), electron-only for n-type  (middle) and hole-only for p-type (right) dopants. The bottom panels depict the corresponding energy levels. $E_g\sim 1$ eV is the bandgap and $E_I \sim 10$ meV is the dopant ionization energy.
\label{fig:Schematic}
}
\end{figure}

We propose to extend the reach of semiconductor detectors towards lower DM masses through a simple innovation: adding shallow impurities to the semiconductor. 
Shallow impurities, also called dopants, are atoms that introduce new energy levels close to the conduction band (n-type dopant) or valence band (p-type dopant), which are populated by electrons or holes contributed by the n- or p-type atoms.
By emitting these charges into the conduction or valence bands, dopants can be ionized with sub-bandgap energy depositions, as schematically shown in Fig.~\ref{fig:Schematic}. Since the dopant's electrons or holes are weakly bound, their orbits lie far from the impurity centers, so the ionization energies are largely independent of the details of the dopant and are instead mostly set by the macroscopic properties of the underlying semiconductor~\cite{PhysRev.97.869,PhysRev.98.915,kohn1957shallow,shklovskii2013electronic}.
In this simplified picture the smallness of the energy required to ionize a dopant can be quantified by accounting for the screening of the impurity potential due to the semiconductor's dielectric function $\epsilon\sim 10$, and for the smallness of the effective electron or hole masses compared to the electron mass in vacuum, typically $m_*/m_e \sim 0.1-1$.
These two factors suppress 
the ionization energy
by roughly $m_* / (m_e \epsilon^2)\sim 10^{-3} -10^{-2}$ with respect to the one of a free atom, resulting in energies of order $10-100$~meV. Such low thresholds allow designing ionization detectors that can be used to probe sub-MeV DM scattering or sub-eV DM absorption. In doped semiconductors, DM can also ionize valence-band electrons, so a doped target  retains the signatures of pure semiconductor  targets for $\gtrsim$ eV energy depositions.

For models of DM scattering off electrons via a light mediator, we find that a doped target with an exposure of 100 g-day could probe the 
entire currently unconstrained sub-MeV mass region for DM produced via freeze-in~\cite{Essig:2011nj,Chu:2011be,Dvorkin:2019zdi}, in the absence of dark counts (DC). For sub-eV dark-photon DM absorbed by electrons, an exposure as small as 1 gram-day with DC at the level of existing undoped detectors could probe currently unconstrained absorption cross sections.

Our proposal has the advantage that it could largely rely on existing technology. Doped semiconductor detectors have been fabricated for decades for infrared (IR) light detection~\cite{rogalski2002infrared,rogalski2019infrared}.  Due to the nature of their applications (\textit{e.g.}, infrared astronomy) the DC requirements on existing detectors are above what is needed for DM detection~\cite{rieke2007infrared}.
However, single-electron sensitivity with small DC has been demonstrated in pure semiconductor detectors that collect charge such as SENSEI ($\sim 450$~DC/g-day)~\cite{SENSEI:2020dpa,SENSEI:2021hcn} or phonons, such as SuperCDMS HVeV~\cite{SuperCDMS:2020ymb}, and they could be further reduced with detector improvements~\cite{SENSEI:2020dpa}. 
Thus, our proposal could be realized by combining the design of existing doped semiconductor detectors with the technologies used to obtain low DC in undoped detectors. 

While in the body of this work we focus on studying ionization signals, for energy depositions below the dopant's ionization threshold, doped targets allow for transitions between the ground and excited electron or hole bound states. Upon relaxation to the ground state, phonons are emitted, which could be potentially detected with future single-phonon detectors. We leave the discussion of phonon signals to an appendix. 

We organize this work as follows. We begin by reviewing the theory of electrons in doped semiconductors, focusing on n-type dopants (the situation for p-type dopants is analogous). We then compute the  energy loss function due to ionization in these materials, obtain DM scattering and absorption rates, and project the detector's reach for a silicon target doped with phosphorus.  In the Supplemental Materials, we discuss detector design, backgrounds, other low-threshold detection technologies, the expected phonon signals, and calculational details. 

\section{Electronics of N-type Doped Semiconductors}
The electronic wavefunctions in a perfect crystal potential are given by Bloch wavefunctions $e^{i \mathbf{k} \cdot  \mathbf{r} }  u_\mathbf{nk}(\mathbf{r})$, where $\mathbf{k}$ labels the crystal momenta, $n$ is a band index, and $u_{n\mathbf{k}}$ are functions with the periodicity of the lattice. 
For donor electrons in n-type semiconductors the spectrum differs from the perfect lattice solution due to the impurities. In this case, the wavefunctions can be expressed as a Bloch-state superposition~\cite{PhysRev.97.869,PhysRev.98.915,kohn1957shallow,shklovskii2013electronic}, 
\begin{equation}
\psi= \frac{1}{\sqrt{V}} \sum_{\mathbf{k},n} A_n(\mathbf{k}) e^{i \mathbf{k} \cdot  \mathbf{r} }  u_{n\mathbf{k}}(\mathbf{r}) \quad ,
\label{eq:exactsol}
\end{equation}
where $V$ is the semiconductor's volume, and the Fourier coefficients $A_n(\mathbf{k})$ are found by solving Schr\"odinger's equation for  $\psi$ in the impurity and lattice potentials. The impurity potential can be approximated by the one of a ionic charge screened by the lattice,
\begin{equation}
U(r)=  -\frac{\alpha}{\epsilon r} \quad ,
\label{eq:impuritypotential}
\end{equation}
where $\alpha$ is the fine-structure constant, and $\epsilon$ is the crystal's dielectric function. The crystal potential is more complex and leads to dependency of the wavefunctions on the band structure. Full knowledge of the bands, however, is not required to obtain approximate wavefunctions, as donor electrons bind only weakly to the impurity so their typical momentum lies close to the bottom of the conduction band. This leads to two simplifications. First, the wavefunctions $u_{n\mathbf{k}}$ in Eq.~\eqref{eq:exactsol} can be approximated to be those at the (possibly degenerate) conduction band minima.
Fixing the band index $n$ to correspond to the conduction band and dropping it, and taking the momentum coordinate of the  $\xi$-th degenerate conduction-band minimum to be $\mathbf{k_\xi}$, this corresponds to approximating $u_{n\mathbf{k}} \approx u_{\mathbf{k_\xi}}$, so the wavefunction Eq.~\eqref{eq:exactsol} for an electron with momentum close to the $\xi$-th minimum simplifies to
\begin{equation}
\psi_\xi \approx e^{i \mathbf{k_\xi \cdot  \mathbf{r} }} u_{\mathbf{k_\xi}}(\mathbf{r})   F(\mathbf{r})  \quad ,
 \label{eq:effmass}
 \end{equation}
where $F(\mathbf{r})\equiv \sum_\mathbf{k} A(\mathbf{k}) e^{i \mathbf{k} \cdot{\mathbf{r}}}/\sqrt{V}$ and $\mathbf{k}$ is now the momentum relative to the band minimum. Eq.~\eqref{eq:effmass} indicates that donor electrons are described by bottom-of-band Bloch wavefunctions modulated by an ``envelope'' $F(\mathbf{r})$, which is the same for all degenerate minima $\xi$. 

The second simplification is that near the band minima, band energies can be approximated by the leading term in a momentum expansion,
\begin{equation}
E(k)=\frac{\mathbf{k}^2}{2m_*} \quad ,
\label{eq:effectivemass}
\end{equation}
where we have assumed isotropy in momentum space for simplicity (``spherical'' band approximation), and $m_*$ is the electron's effective mass.\footnote{The spherical-band approximation is not precise in indirect bandgap semiconductors such as Si and Ge where the bands are anisotropic. However, as discussed in the appendix, treating the bands as spherical will suffice for our purposes.} With these approximations, the envelope functions and energy eigenvalues can be shown to be solutions of the Schr\"odinger equation with a Hamiltonian set by the screened impurity potential Eq.~\eqref{eq:impuritypotential}  and the kinetic term Eq.~\eqref{eq:effectivemass}~\cite{PhysRev.97.869,PhysRev.98.915,kohn1957shallow,shklovskii2013electronic},
\begin{equation}
-\frac{\nabla^2}{2m_*}F(\mathbf{r}) + U(r) F(\mathbf{r})=E\, F(\mathbf{r}) \quad .
\label{eq:schr}
\end{equation}
The solutions to Eq.~\eqref{eq:schr} are Hydrogenic, so the energies (relative to the conduction band), Bohr radius of the bound electrons, and $1s$ ground-state envelope function are 

\begin{equation}
E_n =  -\frac{\alpha^2}{2n^2}  \frac{m_*}{\epsilon^2} \quad ,  \quad a_* = \frac{\epsilon}{\alpha m_*} \quad , \quad F_{1s}(\mathbf{r})=\frac{e^{-r/a_*}}{\sqrt{\pi a_*^3}}\,, 
\label{eq:En} 
\end{equation}
where $n$ is the principal quantum number. For typical values of $m_*/m_e\sim 0.1-1$ and $\epsilon\sim 10$, the ionization energies $ E_I\equiv E_{n=1}$ are of order $10-100$~meV. The Bohr radius $a_*$ of a dopant electron is $\epsilon(m_e/m_*)\sim  10-100$ times larger than typical lattice spacings $a$, so its typical momentum lies near the origin of the first Brillouin zone, $|\mathbf{k}| \ll 1/a$, validating the bottom-of-band approximation. With our approximations both the ionization energy and Bohr radius are independent of the dopant atom, as anticipated in the introduction.

The model described above is the simplest version of the ``effective-mass method''~\cite{PhysRev.97.869,PhysRev.98.915}. 
Corrections to the model arise from wavefunction overlap of electrons localized at different impurities, electron-electron interactions, and differences in the impurity potential at different sites. These corrections lead to dispersion in the discrete energies Eq.~\eqref{eq:En} and thus to ``impurity bands''. For high doping ($n_D\gtrsim 3\cdot10^{18}/\mathrm{cm}^3$ for uncompensated Si:P~\cite{swartz1960low,yamanouchi1967electric,lohneysen1990metal}), these bands allow for electric conduction as in a metal,  resulting in a metal-insulator Mott-Anderson transition~\cite{mott1949basis,anderson1958absence} and eliminating the energy gap that is required for the design of ionization detectors. Here we only consider semiconductors with doping densities below the Mott value. Additional corrections to the effective-mass method come from short-distance modifications to the impurity potential Eq.~\eqref{eq:impuritypotential} that break the degeneracy of the band minima~\cite{aggarwal1965optical,shklovskii2013electronic}. For Si, 
these corrections lead to a ground-state that is set by a superposition of the Bloch wavefunctions at the degenerate minima, modulated by the common 1s envelope of Eq.~\eqref{eq:En} (see appendix).

While up to now we have only considered donor electron bound states, the presence of impurities also affects the conduction-band electron wavefunctions, which are relevant as final states for computing ionization probabilities. The modified conduction-band electron wavefunctions are simply given by Eq.~\eqref{eq:effmass}, with envelope functions set by the positive-energy solutions of the Schr\"odinger Eq.~\eqref{eq:schr}; they are given by~\cite{holt1969matrix}
\begin{eqnarray}
\nonumber
F_{{ \mathbf{k}}}(\mathbf{r})&=& 
\color{black}{\frac{e^{(\frac{\pi}{2ka_*}-i\mathbf{k}\cdot \mathbf{r})} }{\sqrt{V}}}
  \Gamma(1-\frac{i}{ka_*}) 
  {}_1F_1\Big[\frac{i}{ka_*},1,i(kr+\mathbf{k}\cdot \mathbf{r})\Big]  ,
\label{eq:freesol}
\end{eqnarray}
where $\Gamma$ and ${}_1F_1$ are the Gamma and confluent hypergeometric functions.

\section{Dark matter detection using doped semiconductors}

In a doped semiconductor, DM can interact with both valence band and dopant electrons. Since the typical momentum and energy transfer relevant for ionizing dopant electrons 
 are well separated from those for ionizing valence-band electrons, we can treat these two processes independently. The interaction rate on valence-band electrons happens as in pure semiconductors and has been computed in~\cite{Essig:2011nj,Graham:2012su,Lee:2015qva,Essig:2015cda,Derenzo:2016fse,Griffin:2019mvc,Trickle:2019nya,Griffin:2021znd,Hochberg:2021pkt,Knapen:2021run,Knapen:2021bwg,Trickle:2022fwt,An:2014twa,Bloch:2016sjj,Hochberg:2016sqx}. Here we focus instead on obtaining the single-ionization rate of DM interactions with the dopant electrons. Secondary ionization of other dopants by the excited electron are unlikely given the large separation between dopants for doping densities below the Mott transition, so they are not computed here. 
 
 We consider two example DM models. First we take a DM particle $\chi$ scattering with electrons via ultralight vector mediator $A'$ (``dark photon") that kinetically mixes with the SM photon with mixing parameter $\kappa$~\cite{Holdom:1985ag,Galison:1983pa}, so that the momentum-space low energy potential coupling $\chi$ with electrons is
 \begin{equation}
V(\mathbf{q})=\frac{ g_\chi \kappa e}{q^2} \,,
\label{eq:potential}
\end{equation}
 where $g_\chi$ is the coupling between $\chi$ and the mediator, $e$ the elementary charge, and the mediator mass $m_{A'}$ has been neglected with respect to the momentum transfer. Second, we consider a model where the kinetically-mixed vector field itself is the DM, which can be detected by absorption on electrons. 

\begin{figure*}[t!]
\includegraphics[width=0.45\textwidth]{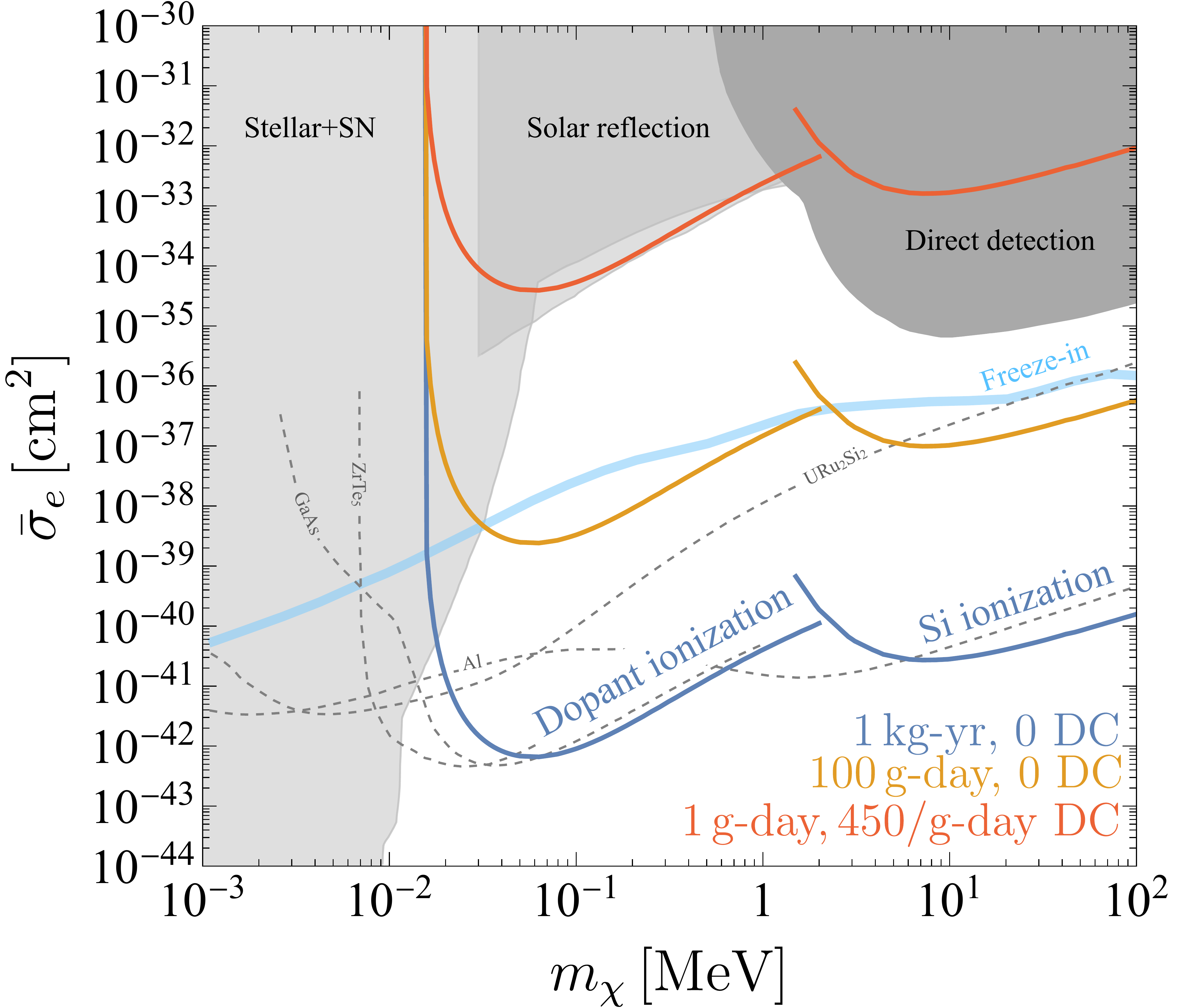}\includegraphics[width=0.45\textwidth]{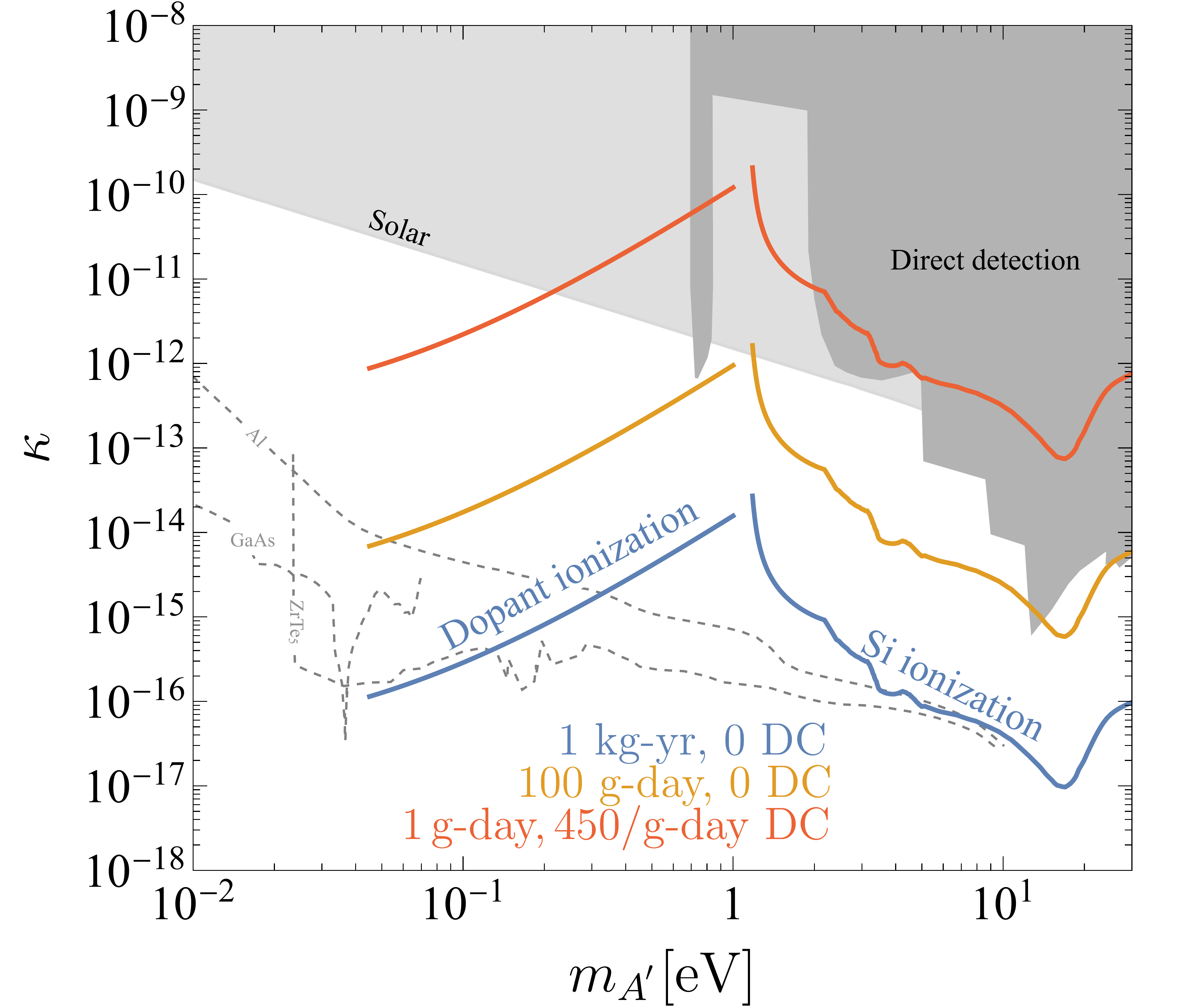}
\caption{\textit{Solid colored lines}: projected $90\%$ C.L. reach  (single-charge ionization signal) for DM-electron scattering via a light mediator (\textbf{left}) and for dark photon DM absorption (\textbf{right}), for a Si:P target with doping density $n_D=1\times 10^{18}\textrm{cm}^{-3}$. Two contributions to the reach are shown, at low and high DM masses coming from P-dopant and Si ionization, respectively. The reach is computed for three levels of exposure and dark counts (DC): 1 kg-yr (blue) and 100 g-day (orange) exposure with zero DC, and 1 g-day exposure (red) with 450 g-day DC. \textit{Shaded grey}: exclusion regions from existing direct detection experiments~\cite{PhysRevLett.125.171802,Chiles:2021gxk, Hochberg:2021yud,PhysRevD.102.042001,PhysRevLett.107.051301}. \textit{Light gray}: excluded by stellar cooling~\cite{Vogel:2013raa}, solar reflection~\cite{An:2017ojc,An:2021qdl} and SN1987A~\cite{Chang:2018rso} (\textbf{left}), or from the solar emission of dark photons~\cite{PhysRevD.102.115022} (\textbf{right}). \textit{Light blue line (left panel):} cross section required for DM being produced by freeze-in ~\cite{Essig:2011nj,Dvorkin:2019zdi}. \textit{Grey-dashed lines:} 
reach of other proposed targets for 1 kg-year exposure and no DC: superconductors~\cite{Hochberg:2015pha} (Al), polar~\cite{Knapen:2017ekk} (GaAs), Dirac~\cite{Hochberg:2017wce} and  Fermi materials~\cite{Hochberg:2021pkt}.
\label{fig:scattering}
}
\end{figure*}

\textbf{Dark matter scattering:} the DM scattering rate per unit target mass is~\cite{Hochberg:2021pkt,Kahn:2021ttr}
\begin{equation}
R = \frac{\rho_{{\chi}}}{\rho_T m_{{\chi}}} \int d^3 \mathbf{\mathbf{v_{\chi}}} f(\mathbf{\mathbf{v_{\chi}}}) \Gamma(\mathbf{v_{\chi}}) \quad ,
\end{equation}
where $\rho_\chi=0.4 \, \mathrm{GeV}/\mathrm{cm}^3$, and $\rho_T$ are the DM and target mass densities (for Si $\rho_T=2.3\, \mathrm{g}/\mathrm{cm}^3$), $f(\mathbf{\mathbf{v_{\chi}}})$ is the DM velocity distribution in the halo~\cite{Drukier:1986tm}, with dispersion, escape, and earth velocities $v_0= 220\,\textrm{km/s}$, $v_{\rm esc}= 500\,\textrm{km/s}$, and $v_E= 240\,\textrm{km/s}$ (in the galactic frame). $ \Gamma(\mathbf{v}_\chi)$ is the scattering rate of a single DM particle with velocity $\mathbf{v_{\chi}}$ in the whole target, given by
\begin{equation}
 \Gamma(\mathbf{v}_\chi)=  \int \frac{d^3 \mathbf{q}}{(2\pi)^3} \left| V(\mathbf{q})\right|^2 \frac{q^2}{2\pi\alpha} \mathcal W(\mathbf{q},\omega_\mathbf{q})\,,
\end{equation}
with $\omega_\mathbf{q}= \mathbf{q}\cdot\mathbf{v_\chi}-q^2/(2m_\chi)$ being the energy transfer, $V(\mathbf{q})$ is given by Eq.~\eqref{eq:potential}, and the energy loss function (ELF) $\mathcal W$ is 
\begin{equation}
\mathcal W(\mathbf{q},\omega) = \frac{(2\pi)^2 \alpha n_D}{{q}^2 |\epsilon(\mathbf{q},\omega)|^2}  \! \sum_{\xi,\mathbf{k}}  \delta(E_\mathbf{k}\!-\!E_{i}\!-\!\omega) \left|\left<\xi \mathbf{k} \left| \mathbf{\hat{\rho}}(\mathbf{q}) \right| i \right>\right|^2 \,,
\label{eq:ELF}
\end{equation}
where $n_D$ the number density of dopants and $\mathbf{\hat{\rho}}(\mathbf{q})=e^{-i \mathbf{q}\cdot \hat{\mathbf{r}}}$ is the momentum-space electron-density operator. In Eq.~\eqref{eq:ELF}, we have already performed the sum over target electrons, so there is a factor of $1/|\epsilon(\mathbf{q},\omega)|^2$ from multi-particle screening~\cite{Trickle:2019nya,Hochberg:2015fth,Kahn:2021ttr}, and the term in brackets is the single-particle form factor between the initial (ground) state and free conduction-band electrons with momentum close to the $\xi$-th minimum. This form factor is obtained in the appendix, where we show that within the effective mass method described previously, and 
for momentum transfers less than the inverse lattice spacing, $|\mathbf{q}|\ll 1/a$, it is given by the form factor between the initial (bound) and free Hydrogenic envelope functions $F_i$ and $F_{\mathbf{k}}$
\begin{equation}
 \sum_{\xi} {\left|\left<\xi \mathbf{k} \left| e^{-i \mathbf{q}\cdot \hat{\mathbf{r}}} \right| i \right>\right|^2} = \bigg|\int d^3\mathbf{r}  \,
 F_{\mathbf{k}}(r)^*   F_{i}(r) e^{-i \mathbf{q}\cdot {\mathbf{r}}} \bigg|^2 \quad .
 \label{eq:hydrogenic}
 \end{equation} 
For ionization from the 1s ground state Eq.~\eqref{eq:En} into the free states Eq.~\eqref{eq:freesol}, Eq.~\eqref{eq:hydrogenic}  has been analytically computed in~\cite{holt1969matrix,Chen:2015pha}. Using the result of~\cite{holt1969matrix} in Eq.~\eqref{eq:ELF} we obtain the ELF for n-type doped semiconductors
\begin{eqnarray}
\label{eq:ELFfinal}
\mathcal W(\mathbf{q},\omega)  &=&\bigg(\frac{E_{\mathrm{eff}}}{E_0} \bigg)^2 \frac{\color{black}{2^{10}\pi^2\alpha m_* n_D a_*^4}}{\color{black}{3 |\epsilon(\mathbf{q},\omega)|^2} } \\ \nonumber
&&  \frac{(3\tilde{q}+\tilde{k}^2+1) \exp\big[-\frac{2}{\tilde{k}} \tan^{-1}\big( \frac{2\tilde{k}}{\tilde{q}^2-\tilde{k}^2+1} \big) \big]}
{[(\tilde{q}+\tilde{k})^2+1]^3 [(\tilde{q}-\tilde{k})^2+1]^3 [1-\exp(-2\pi/\tilde{k})]}
\end{eqnarray}
where $\tilde{q}=q a_*$ and $\tilde{k}\equiv \sqrt{2 m_* (\omega-E_I)} a_*$ is the momentum of the final state electron for an energy transfer $\omega>E_I$. In Eq.~\eqref{eq:ELFfinal}, we have heuristically added a normalization prefactor $({E_{\mathrm{eff}}}/{E_0})^2$ to account for the ratio of local electric field at the donor center and the average field in the crystal, as done in photoionization calculations~\cite{dexter1958theory,anderson1975shallow,sclar1984properties}.

\textbf{Dark matter absorption:} the rate per unit mass for the absorption of kinetically-mixed vector DM is simply $R_{\chi} = \kappa^2 {\rho_{\chi}}/{\rho_T}\,\mathcal{W}(\mathbf{q=0}, m_\chi)$~\cite{Mitridate:2021ctr}, where the ELF is given by Eq.~\eqref{eq:ELFfinal}.

\section{Projected reach}

We now compute the projected DM reach taking for concreteness silicon doped with phosphorus as the target.
For this material, the parameters entering in the ELF Eq.~\eqref{eq:ELFfinal} are set to $E_I=45$~meV,  $m_*=0.3\,m_e$, $a_*=23$~atomic units, and ${E_{\mathrm{eff}}}/{E_0}=2.2$, as discussed in the appendix. The resulting  projections for DM scattering and absorption are presented in Fig.~\ref{fig:scattering}, where in both cases the phosphorus density is $n_D=10^{18}/\mathrm{cm^3}$. This density is chosen to maximize the number of target electrons while staying below the metal-insulator transition. For DM scattering, the bounds are presented as a function of the reference cross section on electrons $\bar{\sigma}_e \equiv { \mu_{\chi e}^2}/{\pi} |V(q=\alpha m_e)|^2$,
where $\mu_{\chi e}$ is the DM-electron reduced mass. For DM absorption the bounds are presented as a function of the photon-dark photon mixing parameter $\kappa$. Bounds are presented for three assumptions regarding dark counts (DC) and exposure. In blue and green we present limits for kg-year and 100 g-day exposures that consider only Poisson statistical uncertainties and assume no DC are observed. In red we project more conservative sensitivities with exposures and DC in line with SENSEI at MINOS~\cite{SENSEI:2020dpa}, that is a 1 g-day exposure and 450/g-day DC. Each projected exclusion curve is broken into two pieces to highlight the reach due to the ionization of dopants (masses below $\sim 1$~MeV or $\sim 1$~eV for scattering or absorption), or due to the ionization of valence band electrons (for larger masses).\footnote{The Si  ionization sensitivity projections include screening, and hence the 1 g-day curve assuming a DC of 450/g-day is weaker than the SENSEI limit shown in gray~\cite{SENSEI:2020dpa}.} 

Our projections show that doped semiconductor targets have a significant discovery potential, and compare favorably against other proposed targets. From the figures we clearly see how introducing doping into the material extends the reach of semiconductor detectors to lower DM masses. In Fig.~\ref{fig:scattering}, we see that in the absence of DC, doped semiconductors have the potential to probe scattering cross sections that lead to DM being produced by freeze-in 
 down to the smallest allowed masses, $m_{\chi}\simeq 30$ keV.  The excellent reach to DM scattering is in part explained by kinematic matching, as for the typical momentum tranfers $q_{\mathrm{typ}}\approx m_{\chi} v_\mathrm{rel}\approx  m_{\chi} v_e \approx 100\,\mathrm{eV} (m_{\chi}/100\,\mathrm{keV})$, with $v_e\approx \alpha/\epsilon\sim 10^{-3}$ ($\approx v_{\chi}$)  being the dopant electron velocity, the energy transfers are precisely of the order of dopant ionization energies,  $q_{\mathrm{typ}} v_{\chi}\approx 100$~meV.  The kinematics are further studied in the appendix, where we present differential scattering rates. The right panel of Fig.~\ref{fig:scattering} indicates that Si:P also has excellent reach to dark-photon absorption for DM masses $m_{A'}\lesssim 1$~eV. For DM absorption, even with a small 1 g-day exposure and if DC are included in our estimates at the level currently observed by SENSEI, our proposal could probe currently unconstrained parameter space.

\section{Discussion}

We have proposed a new DM detection strategy based on looking for DM interactions on dopant atoms. By computing DM interaction rates, we have shown that doped semiconductor targets have the potential to explore large regions of parameter space of two benchmark DM models, sub-MeV DM coupling to the SM via a light mediator, and kinetically-mixed sub-eV dark-photon DM. 
This work begins the exploration of doped sub-MeV DM detectors. From the theoretical side, interaction rates on other doped targets and calculations of the Migdal effect on dopants will be required, while from an experimental perspective developments for the detector design are needed, as discussed in the appendix. We conclude that the development of doped semiconductor detectors with low dark counts is both scientifically and technologically motivated, and may lead to the discovery of DM.

\section{Acknowledgments}

We would like to thank Steve Holland, Junwu Huang, Noah Kurinsky, Guillermo Moroni, Sae Woo Nam, Roger Romani, Miguel Sofo Haro, Javier Tiffenberg, and Sho Uemura for useful discussions. The work of P.D.~is supported by the US Department of Energy under grant DE-SC0010008. D.E.U.~is supported by
Perimeter Institute for Theoretical Physics and by the Simons Foundation. Research at Perimeter Institute is supported
in part by the Government of Canada through the Department of Innovation, Science and
Economic Development Canada and by the Province of Ontario through the Ministry of Economic Development, Job Creation and Trade. R.E.~acknowledges support from DoE Grant DE-SC0009854, Simons Investigator in Physics Award 623940, the US-Israel Binational Science Foundation Grant No.~2016153, and the Heising-Simons Foundation Grant No.~79921. 
M.S.~acknowledges support from Department of Energy Grants DE-SC0009919 and DE-SC0022104.

\newpage
\section{Supplementary Material}

\subsection{Detector design and backgrounds}

To realize the discovery potential of doped semiconductor targets a scalable single-electron detection technology with low dark counts is required. Here we discuss one possible implementation, based on the Skipper-CCD detectors used by the SENSEI experiment~\cite{SENSEI:2020dpa}. Skipper-CCDs are imaging detectors that use high-resistivity n-type Si as the bulk absorber to detect ionization signals. Even if the target is n-type, these CCDs are not doped ionization detectors, since their doping density is extremely small, on the order of $10^{11}/\textrm{cm}^3$, and since they are operated at temperatures where the dopants are already ionized. Skipper-CCDs have demonstrated single electron-hole-pair resolution and dark counts as low as $450$ events per gram-day~\cite{SENSEI:2020dpa}. Building upon this technology, we envision designing a Skipper-CCD with a large level of n-type doping in the bulk.\footnote{A detector that instead measures ionization signals by reading out phonons such as SuperCDMS HVeV could also be considered, but the current single-electron dark current in such detectors is larger than at SENSEI by a factor of $10^3$~\cite{SuperCDMS:2020ymb}. In addition, SuperCDMS HVeV, at least in its current setup, would reject a single-electron or single-hole signal from an ionized dopant, as its single-electron bin corresponds only to events that are electron-hole pairs.}

The design of a doped Skipper-CCD would require several technological developments. For concreteness, we discuss these developments in the context of n-type targets (the situation for p-type detectors is analogous). First and foremost, current n-type Skipper-CCDs used for detecting DM only collect \textit{holes}, and they do so in a ``buried channel'' located right below the detector's frontside where charges are stored until readout. For our proposal to be realized, a doped n-type Skipper-CCD needs to collect  ionized electrons from n-type dopants. A detailed design that satisfies this requirement will be presented in future work. 

In order to collect the charge signals, and to ensure low levels of charge trapping, the detector would need to be placed in an electric field (as a conventional CCD), which would be provided by a bias voltage applied to detector contacts. Since the active area would be doped, this electric field would also induce currents within  the ``impurity band'' formed by the dopant's energy levels, without the need for electrons being excited into the conduction band (``hopping conductivity''~\cite{PhysRev.79.726}), which would constitute a dark current. A technology to eliminate this current exists, which is used in a type of doped semiconductor detectors called ``Blocked Impurity Band'' detectors~\cite{petroff1986blocked} (BIBs). The idea is to introduce a layer of undoped semiconductor between the active detection region and the charge readout stages, which in our case are the buried channels, so that impurity band conduction is blocked. BIBs that operate with high levels of doping have quantum efficiencies of the order of $80\%$ for $\sim 1$~V of applied voltage across an absorber that is tens of $~\mu$m wide~\cite{rogalski2019infrared}, which also indicates that charge trapping in these devices is under control. This also suggests that charge transfer across pixels (which is required for readout in a CCD-like setup) could be done efficiently even if the target is doped. 

In order to suppress thermal generation of dark currents due to the ionization of dopants, the detector would need to be operated at cryogenic temperatures. The rate for the thermal generation of electrons from neutral donors can be estimated to be $\langle\sigma_{eD^+}v_e\rangle N_c e^{-E_I/T}$ from detailed balance~\cite{MARTINI1972181}. Here $N_c$ is the effective density of states of electrons in the conduction band, $\sigma_{eD^+}$ is the cross section for electrons capture by charged donors, and $\langle\rangle$ indicates the thermal average. From measurements, we obtain $\langle\sigma_{eD^+}v_e\rangle\approx 7\times10^{-6} \textrm{cm}^3/\textrm{s}$ for Si:P~\cite{PhysRev.144.781}. Therefore, thermal dark currents would be kept at the level of $1\, e^-/\mathrm{kg}/\mathrm{yr}$ for $n_{D}=10^{18}/\textrm{cm}^3$ by operating the detector at a temperature $T\approx 5$ K ($\approx 0.43\,$meV). 

The low operating temperatures represent a challenge for a doped detector based on a CCD design, since conventional CCDs cannot operate at temperatures below $\sim 70$ K due to carrier freeze-out ~\cite{janesick1987scientific}, where the gates become non-conducting. This issue could be solved by replacing the standard polysilicon gates with metal gates,  as done in some metal oxide semiconductor (MOS) devices. 

In addition to thermally generated dark counts, it is likely that a doped Skipper-CCD would be affected by the backgrounds that are observed in undoped Skipper-CCDs. The origin of the currently observed backgrounds in Skipper-CCDs is unknown, even if a fraction of them have been shown to arise from secondary radiation of high energy tracks~\cite{Du:2020ldo,SENSEI-radiative-to-appear}.  Track-induced ``extrinsic'' backgrounds can be reduced by working in a radio-pure environment and improving shielding. The ``intrinsic'' detector backgrounds in Skipper-CCDs, on the other hand, could arise from charge leakage into the detector contacts, or from slow release of electrons from unidentified traps. Both of these effects could in principle be reduced by improving the insulating layers and pre-filling empty traps. On the other hand, the intrinsic dark counts that are observed in doped blocked-impurity band detectors are possibly coming from conduction across the blocking layers into the contacts~\cite{wang2016analysis,marozas2018surface,pan2021dark}, either via tunneling or due to impurities. Such dark currents could be suppressed by increasing the thickness of the insulating layers~\cite{wang2016analysis}. Yet another source of dark counts in doped detectors arises due to conduction within the impurity band by hopping into neutral (occupied) dopants, a process that's referred to as $\epsilon_2$ hopping conductivity~\cite{1971JETPL..14..185G,shklovskii2013electronic}. This can be exponentially suppressed  by reducing the doping density, but measurements are required to find an optimal doping value. 

We conclude this section by pointing out that several other detector targets with sub-eV thresholds exist, beyond doped semiconductors, which are used to detect infrared (IR) radiation but that can also be used to detect sub-MeV DM scattering or sub-eV dark photon absorption. We provide an incomplete list of available IR detectors in Table~\ref{tab:IR_detectors}, which includes Superconducting nanowire detectors (SNPDs), Single-photon avalanche diodes (SPADs), Mercury-Cadmium-Telluride detectors (HgCdTe), the already mentioned BIBs, Quantum Dot and Quantum Well Infrared Photodetectors (QDIP and QWIP) and Transition Edge Sensors (TES). These detectors can be designed and used as targets for DM absorption or scattering, or in some cases can be coupled to read out excitations from an external absorber that acts as the target. 

SNSPDs were proposed as sub-MeV dark-matter targets in~\cite{Hochberg:2019cyy}, but an $\sim 8$ order-of magnitude increase in detector exposure (without a corresponding increase in dark currents) would be needed to probe regions of parameter space that are not currently excluded by astrophysical searches~\cite{Hochberg:2021yud}. Current nanogram-scale SNSPD detectors have dark counts of the order of $10^{-6}$ Hz. The origin of dark counts in SNSPDs is currently unknown, but it is possible that they arise from secondary emission from environmental high-energy radiation~\cite{Du:2020ldo}, or due to micro-fractures upon detector cooling~\cite{Anthony-Petersen:2022ujw}. Both of these dark count sources increase with detector exposures. SNSPDs have also been used as detectors to measure interaction events of DM on an external dielectric stack target~\cite{Baryakhtar:2018doz,Chiles:2021gxk}, and have also been proposed as photodetectors, to detect the photons that arise from DM with masses $\gtrsim$1~MeV interacting in a nearby scintillating target~\cite{Derenzo:2016fse,Essig:2019kfe,Blanco:2019lrf,Blanco:2022cel}.

Another possibility is to consider detectors based on low-bandgap compounds of the III-V groups. One of the most mature single-photon detection technologies where these compounds have been used are single-photon avalanche diodes, or SPADs. These detectors have the advantage that they be operated at significantly higher temperatures than SNSPDs, but suffer from larger dark currents~\cite{dello2022advances}, likely due to tunneling~\cite{huang2020high}. A type of SPADs (``Silicon Photomultipliers'' or SIPMs) have been proposed to study DM scattering with an energy threshold of $150$\,eV \cite{SIPM}, but to our knowledge the potential of SPADs as sub-MeV DM detectors has not been explored.

\begin{table}[t]
 \begin{tabular}{|c|c|}
 \hline
{Detector type} &{Energy gap (eV)}    \\
 \hline
 SNSPD~\cite{Chiles:2021gxk} & $ 10^{-3}$    \\
 \hline
  III-V SPAD~\cite{zhang2015advances} & $ 0.1$   \\
 \hline
\multirow{2}{*}{Hg$_{1-x}$Cd$_{x}$Te~\cite{doi:10.1146/annurev.astro.44.051905.092436}}  & 0.5 ($x=0.44$)   \\
&0.1 ($x=0.194$) \\
\hline
Doped semiconductor& $5 \times 10^{-2}$ (Si:P, Si:As) \\
 BIB~\cite{doi:10.1146/annurev.astro.44.051905.092436} &$10^{-2}$ (Ge:Ga) \\
&$6\times 10^{-3}$(GaAs:Te) \\
 \hline
QDIP~\cite{Campbell20071815}&$ 0.1$ (InAs/InGaAs)  \\ 
 \hline
QWIP~\cite{ROGALSKI1997295} &$ 0.1$ (GaAs/AlGaAs) \\
\hline
TES~\cite{ROGALSKI1997295} &$  10^{-3}$ \\
 \hline 
\end{tabular}
\caption{Examples of existing IR photon detectors and the approximate energy gaps required to create a measurable target excitation.  
}
\label{tab:IR_detectors}
\end{table}

Yet another option is to consider detectors based on Mercury-Cadmium-Telluride (HgCdTe), a compound that has the advantage that its bandgap can be tuned, in principle down to the metallic transition~\cite{rogalski2019infrared}. HgCdTe detectors are widely used in astronomy for near and mid-infrared detection, but to the best of our knowledge no HgCdTe-based detector has demonstrated single-photon detection, which would be a requirement for searching for sub-MeV DM scattering or sub-eV DM absorption. Significant progress, however, is being made to achieve IR single-photon detection using photodiodes based in HgCdTe~\cite{dello2022advances}. 

Going towards the deep infrared, doped semiconductor BIBs offer some of the leading sensitivities.  That being said and as already discussed in the body of this work, existing BIBs suffer from dark counts that are too large for the purposes of detecting DM. 

Other more recently developed technologies are quantum well and quantum dot detectors. Regarding these detectors, here we only point out that quantum dots have been proposed as targets for DM scattering~\cite{Blanco:2022cel}, but to our knowledge their potential as photodetectors detectors to search for photons from DM interactions in a nearby target has not been explored.

Finally, TES have been proposed as sub-MeV DM detectors in a setup where DM interacts with an external absorber creating phonons, which are then read out by TES located in the absorber's surface~\cite{Hochberg:2015fth,Hochberg:2015pha,Hochberg:2016ajh,SPICE}, but order-of-magnitude improvements in the TES's energy thresholds are required to probe such light DM models~\cite{ren2021design}. Like SNSPDs, TES have also been proposed as a photodetector to detect photons from DM with masses $\gtrsim$1~MeV interacting in a nearby scintillating target~\cite{Derenzo:2016fse,Essig:2019kfe,Blanco:2019lrf,Blanco:2022cel}.  Dark currents for TES's are briefly discussed in the section ``Sub-ionization energy depositions and phonon signals''.

\subsection{Phosphorus-doped Silicon ELF parameters}

In this appendix, we discuss our choices for the numerical values of the parameters entering into the ELF Eq.~\eqref{eq:ELFfinal} for the specific case of silicon doped with phosphorus, Si:P.
First, we set the ionization energy $E_I$ to the experimentally measured value, $E_I=45$~meV~\cite{aggarwal1965optical,ning1971multivalley}. For the Coulomb impurity potential and spherical band approximations used to obtain the ELF, 
Eq.~\eqref{eq:En} sets the relations that allow the effective mass and Bohr radius to be calculated from the ionization energy. These relations (and more generally the effective mass method used in this work) must be regarded only as a rough approximation for Si,  as for this material the band structure around the minimum is anisotropic, the impurity potential deviates from the Coulomb form near the impurity ion, and intervalley couplings modify the spectrum~\cite{aggarwal1965optical,ning1971multivalley}. In order to account for these corrections, we use the phenomenological prescription of~\cite{ning1971multivalley}, which has been demonstrated to correctly describe the experimentally measured energy levels in Si:P. The prescription retains the spherical band approximation and the Hydrogenic form of the envelope functions (so we may retain the form of the ELF Eq.~\eqref{eq:ELFfinal}), but treats $m_*$ as a free parameter that is determined by matching the ground state energy levels obtained with the spheroidal and realistic ellipsoidal band case in Si,
and $a_*$ as a variational parameter that is chosen to minimize the full Hamiltonian including short-distance and intervalley corrections. This results in $m_*=0.3\,m_e$ and $a_*=23$~atomic units~\cite{ning1971multivalley}. 
For the screening prefactor in the denominator of Eq.~\eqref{eq:ELFfinal}, $1/|\epsilon(\mathbf{q},\omega)|^2$, we simply approximate $|\epsilon(\mathbf{q},\omega)|\approx|\epsilon(0,0)|$, which leads to a small error of order $O((q/\alpha m_e)^2)$ in the calculations~\cite{doi:10.1143/JPSJ.20.778,doi:10.1143/JPSJ.21.1852}.
Finally, we set the normalization prefactor to ${E_{\mathrm{eff}}}/{E_0}=2.2$ to match the normalization of measurements of the dielectric function in Si:P at $\mathbf{q}=0$ (\textit{i.e.}, photoabsorption measurements) reported in~\cite{thomas1981optical,gaymann1995temperature}, by using the relation
\begin{equation}
\mathcal W(\mathbf{q},\omega)  = \mathrm{Im}\bigg[-\frac{1}{\epsilon(\mathbf{q},\omega)}\bigg] \quad ,
\label{eq:elfrelation}
\end{equation}  
evaluated at $\mathbf{q}=0$. In order to validate our calculation of the ELF, we have compared our results with the frequency-dependent photoabsorption data of~\cite{thomas1981optical,gaymann1995temperature} and confirmed that it correctly reproduces the data (see section below on ELF validation). Note that by using Eq.~\eqref{eq:elfrelation}, it is in principle possible to obtain the ELF directly from measurements of the material's dielectric constant instead of using our computation, Eq.~\eqref{eq:ELFfinal}. To our knowledge, however, no measurements of $\epsilon(\mathbf{q},\omega)$ in Si:P away from $\mathbf{q}= 0$ have been performed,
and even at $\mathbf{q}=0$ the dielectric function has only been measured over a limited range of frequencies, so in what follows we rely on Eq.~\eqref{eq:ELFfinal} for obtaining DM interaction rates and calculating projections.

\subsection{Form factor derivation}
In this section, we calculate the form factor for dopant electron transitions between the ground state and free conduction-band energy levels.  We begin by defining Bloch wavefunctions for conduction-band electrons  as
\begin{equation}
\phi_{\xi \mathbf{k}}=\frac{1}{\sqrt{V}}e^{i \mathbf{k} \cdot  \mathbf{r} } e^{i \mathbf{k}_\xi \cdot  \mathbf{r} }   u_\mathbf{k}(\mathbf{r}) \,,
\end{equation}
where the index $\xi$ specifies one of the possibly degenerate conduction-band minima and $\mathbf{k}_\xi$ its position in momentum space, $\mathbf{k}$ labels crystal momenta measured from this minimum, and $V$ is the semiconductor's volume. The periodic functions $u_\mathbf{k}$ are set to be unit-normalized in $V$, 
\begin{equation}
\int_{V} d^3\mathbf{r}  |u_\mathbf{k}(\mathbf{r})|^2 =1  \,,
\end{equation}
and Bloch wave-functions are orthonormal, 
\begin{equation}
\left<\phi_{\eta \mathbf{k'}}|\phi_{\xi \mathbf{k}}\right>=\delta_{\xi \eta} \delta_\mathbf{k-k'} \quad .
\end{equation}
As discussed in the section Electronics of Doped Semiconductors, the initial (ground) state and final single-particle states are given by a sum of conduction-band Bloch wavefunctions,
\begin{eqnarray}
\left|  i \right> &=& \sum_{\eta,\mathbf{k}} \alpha_{\eta}   A_i(\mathbf{k}) \left| \phi_{\eta \mathbf{k}} \right> \quad , \\
\left|  f \right> &=& \sum_{\eta',\mathbf{k'}}  \beta_{\eta'}    A_f(\mathbf{k}') \left| \phi_{\eta' \mathbf{k'}} \right> \quad ,
\end{eqnarray}
where the coefficients setting the components of the wavefunctions at the different minima are normalized as 
\begin{equation}
\sum_\eta |\alpha_\eta|^2 =1 \quad .
\label{eq:norm2} \quad\end{equation}
As an example, for the Si ground state, which is relevant for our limit projections, the initial ground state is a singlet s-wave state formed by equal-weight superpositions of the wavefunctions at these minima, modulated by the 1s envelope function of Eq. \eqref{eq:En},
\begin{equation}
\psi_{1s}^{\mathrm{Si}}=F_{1s}(\mathbf{r}) \sum_{\xi=1}^{6}\,\frac{1}{\sqrt{6}}e^{i \mathbf{k}_\xi \cdot  \mathbf{r} }  u(\mathbf{k_\xi},\mathbf{r})  \quad ,
\label{eq:initialstate}
\end{equation}
so for Si, $\alpha_{\eta}=1/\sqrt{6}, \eta=1,2,..., 6$.

Now, the form factor for transitions between the initial and final states is given by
\begin{eqnarray}
\nonumber 
{\left|\left<f \left| \hat{\rho}(\mathbf{q}) \right| i \right> \right|^2} &=&\bigg| \sum_{\eta \eta'\mathbf{k}\mathbf{k'}} \beta_{\eta'}^* \alpha_{\eta} 
\\ \nonumber &&A_f^*(\mathbf{k'})  A_i(\mathbf{k}) \left<\phi_{\eta' \mathbf{k'}} \left| e^{-i \mathbf{q}\cdot \hat{\mathbf{r}}} \right| \phi_{\eta \mathbf{k}} \right> \bigg|^2 \quad , \\
\label{eq:form1}
\end{eqnarray}
where $\hat{\rho}(\mathbf{q})= e^{-i \mathbf{q}\cdot \hat{\mathbf{r}}} $ is the momentum-space electron density operator. 
Given that the different conduction band minima in typical semiconductors such as Si and Ge are separated by distances of order $1/a\sim 1$~keV in momentum space, with $a$ being the lattice spacing, for momentum transfers $q \ll 1$~keV the generator of momentum translations $e^{-i \mathbf{q}\cdot \hat{\mathbf{r}}} $ does not connect different conduction band minima, and we obtain the simplification
\begin{equation}
 \left<\phi_{\eta' \mathbf{k'}} \left|e^{-i \mathbf{q}\cdot \hat{\mathbf{r}}} \right| \phi_{\eta \mathbf{k}} \right> = \delta_{\eta' \eta} \delta_{\mathbf{k'}-\mathbf{k}-\mathbf{q}} \quad .
 \label{eq:simpl1}
\end{equation}
This simplification is valid for sub-eV absorption in the semiconductor, where $q\lesssim 1$~eV, and for sub-MeV DM scattering via a light mediator, where we have checked that the typical momentum transfer is  $q_{\mathrm{typ}}\approx m_{\chi} v_\mathrm{rel}\approx  m_{\chi} v_e \approx 100\,\mathrm{eV} (m_{\chi}/100\,\mathrm{keV})$, with $v_e\approx \alpha/\epsilon\sim 10^{-3}$. Inserting Eq.~\eqref{eq:simpl1} in Eq.~\eqref{eq:form1} we obtain
\begin{eqnarray}
\nonumber 
{\left|\left<f \left| \hat{\rho}(\mathbf{q}) \right| i \right> \right|^2} &=& \bigg| \sum_{\eta\mathbf{k}} \beta_{\eta}^* \alpha_{\eta} \,
 A_f^*(\mathbf{k+q})  A_i(\mathbf{k}) \bigg|^2 \quad ,\\
 \label{eq:form2}
\end{eqnarray}
and using Parseval's theorem on Eq.~\eqref{eq:form2}, we get
\begin{eqnarray}
\nonumber 
{\left|\left<f \left| \hat{\rho}(\mathbf{q}) \right| i \right> \right|^2} &=& \bigg|\sum_{\eta}\beta_{\eta}^* \alpha_{\eta} \bigg|^2  \bigg| \int d^3\mathbf{r}  \,
 F_f(\mathbf{r})^*   F_i(\mathbf{r}) e^{-i \mathbf{q}\cdot {\mathbf{r}}} \bigg|^2\,,\\
 \label{eq:form3}
\end{eqnarray}
where we defined Fourier transforms as $F(\mathbf{r})\equiv \sum_\mathbf{k} A(\mathbf{k}) e^{i \mathbf{k_\xi}}/\sqrt{V}$  that are unit-normalized in the target volume. Now, take the final state to be $\left|f\right>=\left |\xi \mathbf{k}\right>$ so $\beta_\eta = \delta_{\eta \xi}$. Then Eq.~\eqref{eq:form3} simplifies to
\begin{eqnarray}
\nonumber 
{\left|\left<\xi \mathbf{k} \left| \hat{\rho}(\mathbf{q}) \right| i \right> \right|^2} &=& | \alpha_{\xi} |^2  \bigg| \int d^3\mathbf{r}  \,
 F_{\mathbf{k}}(\mathbf{r})^*   F_i(\mathbf{r}) e^{-i \mathbf{q}\cdot {\mathbf{r}}} \bigg|^2 \quad .\\
\end{eqnarray}
Finally, summing over the final-state minima $\xi$ and using the normalization condition Eq.~\eqref{eq:norm2} we obtain
\begin{eqnarray}
\nonumber 
 \sum_{\xi} {\left|\left<\xi \mathbf{k} \left| \hat{\rho}(\mathbf{q}) \right| i \right> \right|^2} &=& \bigg|\int d^3\mathbf{r}  \,
 F_{\mathbf{k}}(r)^*   F_{i}(r) e^{-i \mathbf{q}\cdot {\mathbf{r}}} \bigg|^2 \quad ,
 \label{eq:finalresult}\\
\end{eqnarray}
which is the result quoted in Eq.~\eqref{eq:hydrogenic}.

\subsection{ELF validation and comparison}
\begin{figure}[t!]
\includegraphics[width=0.45\textwidth]{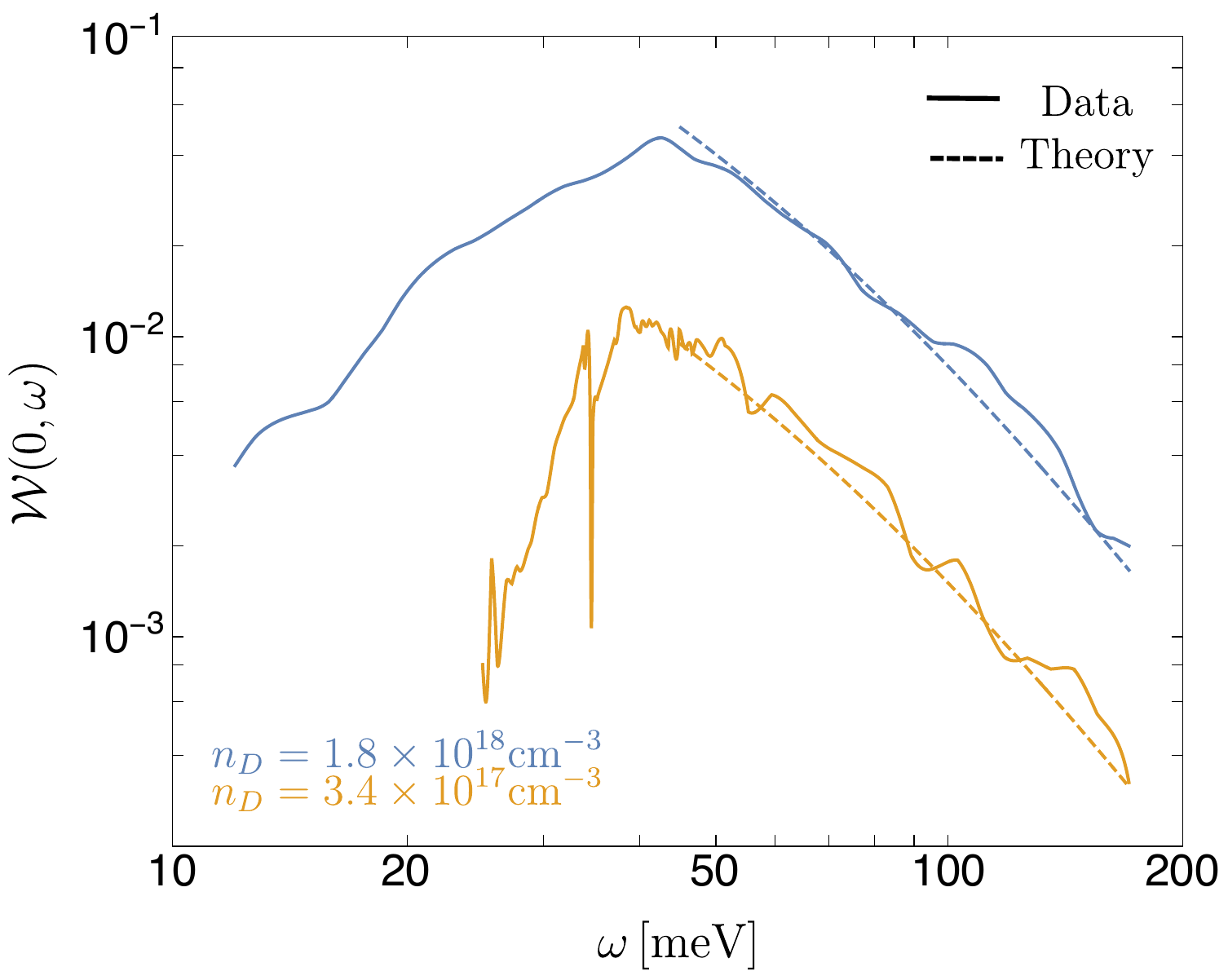}

\caption{The ELF in the optical limit $\mathcal W(0,\omega)$ as a function of the photon energy $\omega$ for Si:P at 10~K. The solid lines denote the ELF derived from the measurement of optical reflectance with $n_D=1.8\times 10^{18}\textrm{cm}^{-3}$ (blue) and $n_D=3.4\times 10^{17}\textrm{cm}^{-3}$ (orange)~\cite{PhysRevB.52.16486,PhysRevLett.71.3681}.  The dashed lines show the analytical results based on the hydrogenic ELF shown in Eq.~(\ref{eq:ELFfinal}). 
\label{fig:ELF_validation}
}
\end{figure}

In this section, we validate our analytical approximation of the ELF in Si:P (Eq.(\ref{eq:ELFfinal})) by comparing it to experimental data and alternative analytic ELF computations. To our knowledge, only optical data is available for Si:P, meaning only $\mathcal W(0,\omega)$ is determined by experiments (within a range of measured frequencies).
 In Fig.~\ref{fig:ELF_validation}, we show $\mathcal W(0,\omega)$ for Si:P calculated from the measurement of optical reflectance with two doping densities~\cite{PhysRevB.52.16486,PhysRevLett.71.3681}, and we also present our analytical hydrogenic ELF Eq.~(\ref{eq:ELFfinal}). For both doping densities, the hydrogenic ELF matches the data very well for energies above the ionization threshold $E_I=45$ meV.

To check the robustness of the hydrogenic ELF at finite $\mathbf{q}$, we compare it with another analytical expression for the ELF proposed by Mermin, where an empirical formula of the dielectric function, or equivalently the ELF, is provided with coefficients to be fitted from data. The Mermin ELF relevant for dopants in Si has the following form:
\begin{eqnarray}\label{eq:Mermin_ELF}
\mathcal W_{\rm Mer}(\mathbf{q},\omega)=A\,\textrm{Im}\left[\frac{-1}{\epsilon_{\rm Mer}(\mathbf{q},\omega)}\right],
\end{eqnarray}
where $\epsilon_{\rm Mer}$ is the Mermin dielectric function defined as~\cite{Mermin:1970zz}
\begin{eqnarray}\label{eq:Mermin_epsilon}
\epsilon_{\rm Mer}(\mathbf{q},\omega)=1+\frac{(1+i\Gamma/\omega)(\epsilon_{\rm Lin}(q,\omega+i\Gamma)-1)}{1+(i\Gamma/\omega)\frac{\epsilon_{\rm Lin}(q,\omega+i\Gamma)-1}{\epsilon_{\rm Lin}(q,0)-1}}.
\end{eqnarray}
Here $\epsilon_{\rm Lin}(q,\omega)$ is Lindhard dielectric function derived from the free electron gas~
\begin{eqnarray}
\epsilon_{\rm Lin}(q,\omega)=1+\frac{3\omega_p^2}{q^2v_F^2}\lim_{\Gamma\to0}f\left(\frac{\omega+i\Gamma}{qv_F},\frac{q}{2m_e v_F}\right),
\end{eqnarray}
with $v_F=\left(\frac{3\pi\omega_p^2}{4\alpha m_e^2}\right)^{1/3}$ and
\begin{eqnarray}
f(u,z)&=&\frac{1}{2}+\frac{1}{8z}[g(z-u)+g(z+u)]\nonumber\\
g(x)&=&(1-x^2)\textrm{log}\left(\frac{1+x}{1-x}\right).
\end{eqnarray}
Taking the $q\to 0$ limit of Eq.~(\ref{eq:Mermin_ELF}), one gets
\begin{eqnarray}\label{eq:Mermin_ELF_k_0}
\mathcal W_{\rm Mer}(\mathbf{q},\omega)\bigg|_{\mathbf{q}\to 0}=A\, \textrm{Im}\left[\frac{-1}{1-\omega_p^2/(\omega^2+i\Gamma\omega)}\right].
\end{eqnarray}
In Eq.~(\ref{eq:Mermin_ELF_k_0}), $A$, $\omega_p$, and $\Gamma$ are fitting parameters that are determined by matching to data. After fitting to optical data for Si:P with $n_{D}=1.8\times 10^{18}\textrm{cm}^{-3}$ above the ionization energy $E_I=45$\,meV, we get $A=0.065$, $\omega_p=45$\,meV, and $\Gamma=70$\,meV.

\begin{figure}[t!]
\centering
\includegraphics[width=0.45\textwidth]{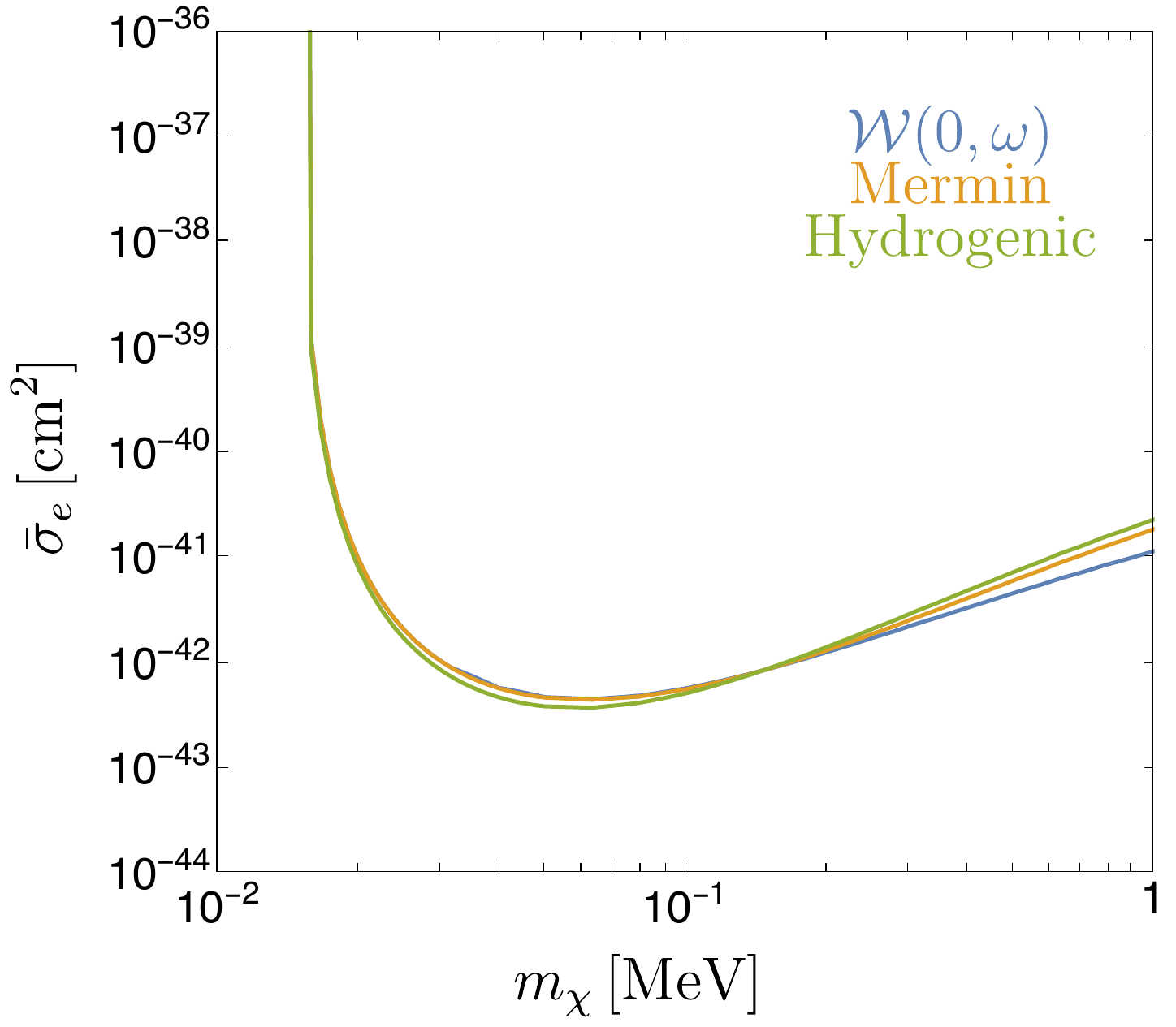}
\caption{ Projected $90\%$ C.L. reach of DM-electron scattering in doped semiconductors (Si:P with $n_D=1.8\times 10^{18}/\textrm{cm}^{-3}$) with a light dark photon mediator assuming 1~kg-yr exposure and zero background. The blue line shows the results using $\mathcal W(0,\omega)$ obtained purely from optical data. The orange and green lines are calculated based on ELF from Mermin ELF (Eq.~(\ref{eq:Mermin_ELF})) and hydrogenic model (Eq.~(\ref{eq:ELFfinal})), respectively.
\label{fig:rate_comparison}
}
\end{figure}

\begin{figure*}[t!]
\centering
\includegraphics[width=1\textwidth]{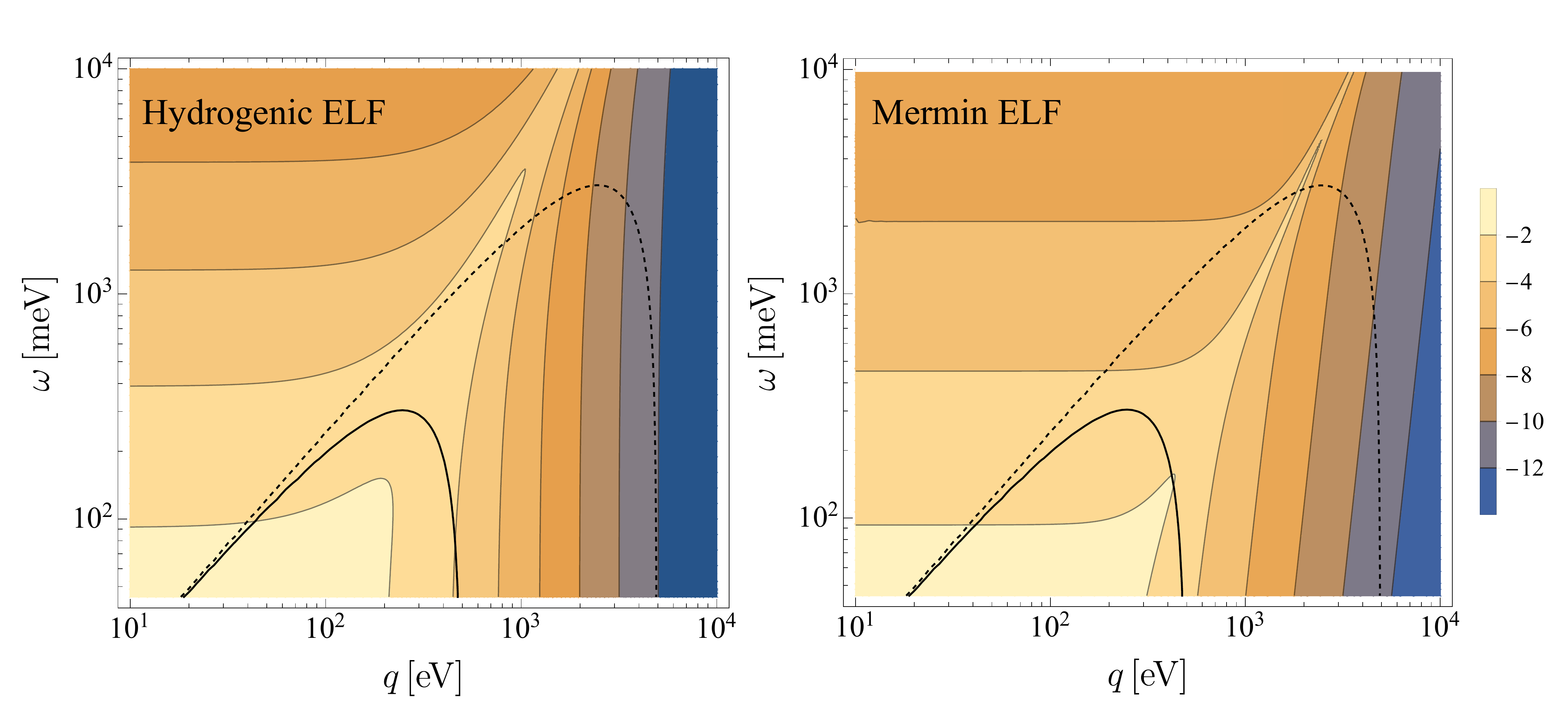}
\caption{Contour plot of the hydrogenic ELF (left) and Mermin ELF (right) as a function of $q$ and $\omega$. The legend shows the value of $\textrm{Log}_{10}[\mathcal W(q,\omega)]$. The black lines show the relation $\omega=q\, (v_{\rm esc}+v_E)-q^2/(2m_{\chi})$ for $m_{\chi}=0.1$ MeV (solid) and $m_{\chi}=1$ MeV (dashed), where $v_{\rm esc}$ is the escape velocity and $v_E$ is the Earth velocity in our galactic frame. The region below the black line is kinematically allowed for a given $m_{\chi}$. 
\label{fig:ELFs}
}
\end{figure*}

We then evaluate the impact of the different ELFs on our computations in Fig.~\ref{fig:rate_comparison}, where we compare the DM rate for dopant ionization in the same material as above (Si:P with $n_{D}=1.8\times 10^{18}\textrm{cm}^{-3}$), calculated from the Hydrogenic and Mermin ELFs. In the figure we have assumed that scattering occurs via a light dark photon mediator. Our results indicate that the rates obtained from the Hydrogenic and Mermin ELFs agree well with each other in the whole sub-MeV region of DM masses. The two approaches are further compared in Fig.~\ref{fig:ELFs}, where we show contours of the ELFs as a function of momentum and frequency. From the figure we observe that at low momentum transfers compared with the characteristic momentum scale of the ELFs, $q\lesssim 1/a_*\sim 100\, \textrm{eV}$, both ELFs are similar, and start to differ only at larger momentum transfers. This feature explains in part the broad agreement of the rates obtained using the two ELFs, as for sub-MeV DM scattering via a light mediator the typical momentum transfer is precisely $q\lesssim 100\, \textrm{eV}$, except for masses near an MeV. Nevertheless, the momentum-dependency of the ELF is relevant at large DM masses. This is shown in Fig.~\ref{fig:rate_comparison}, where we also present in blue the rate obtained from $\mathcal W(0,\omega)$ purely from an interpolation of optical data. We see that this approach starts to differ from the momentum-dependent Hydrogenic and Mermin ELFs for $m_\chi$ near 1~MeV, which indicates that momentum-dependent terms in the ELFs become relevant for such masses. 

\subsection{Differential rate of DM scattering on dopant electrons}

\begin{figure}[t!]
\centering
\includegraphics[width=0.45\textwidth]{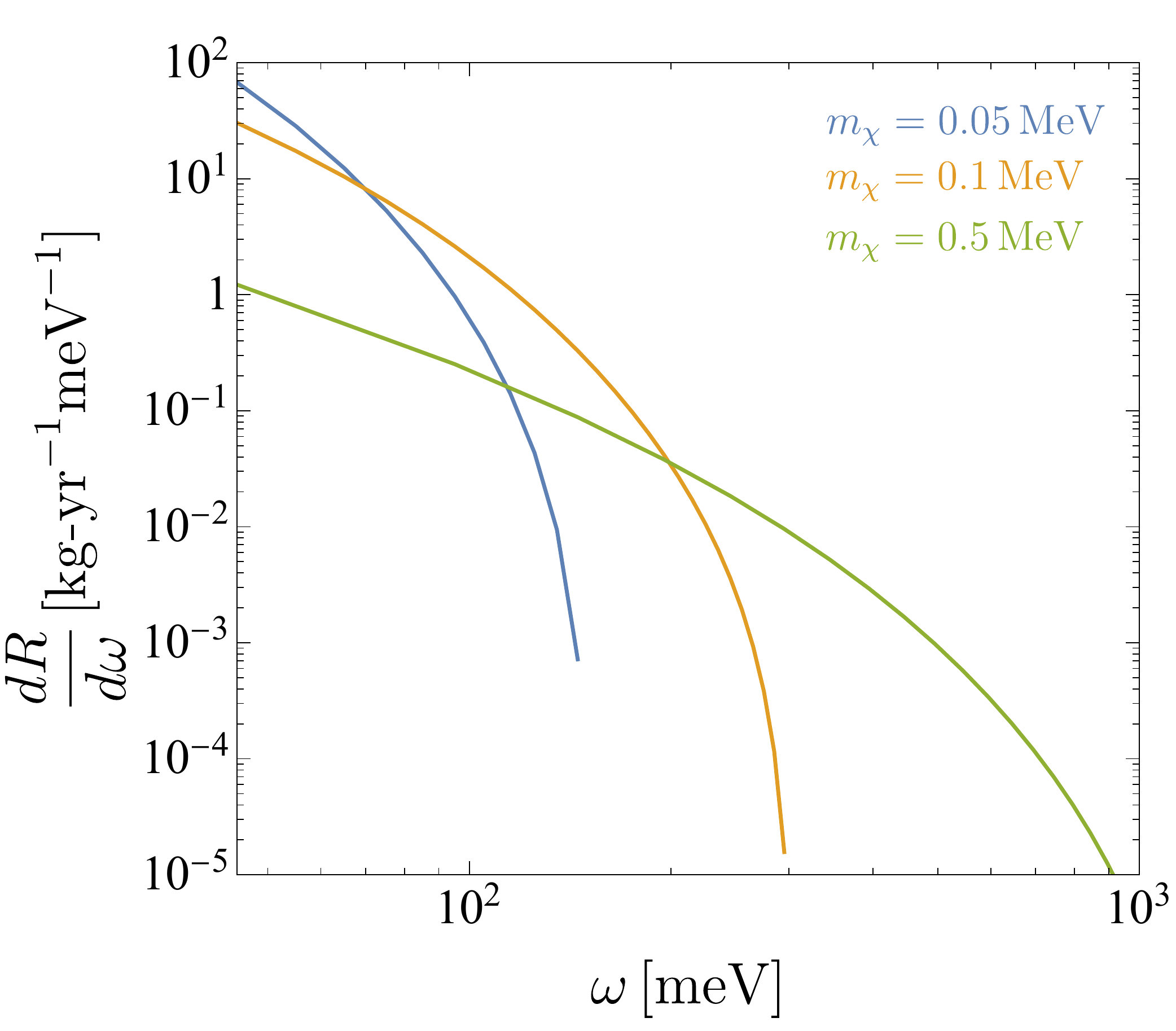}
\caption{The differential rate $\frac{dR}{d\omega}$ for DM ionizing a single dopant electron in Si:P with a light dark photon mediator. Here we choose $n_D=1.8\times 10^{18}/\textrm{cm}^3$ and $\bar\sigma_e=10^{-40}\,\textrm{cm}^2$. The solid lines show the results for three different DM masses: $m_\chi=0.05\,\textrm{MeV}$ (blue), $m_\chi=0.1\,\textrm{MeV}$ (orange), and $m_\chi=0.5\,\textrm{MeV}$ (green).  
\label{fig:energyspectrum}
}
\end{figure}

In this section, we study the differential rate $\frac{d R}{d\omega}$ for DM ionizing dopant electrons. The general $\frac{d R}{d\omega}$ for DM scattering with electrons in a target is given as:
\begin{equation}
\frac{d R}{d\omega}=\frac{\rho_\chi}{(2\pi)^3\alpha\rho_{T}m_\chi}\int dq\,q^3\eta(v_{\rm min}(q,\omega)) \left| V(q)\right|^2 \mathcal W(q,\omega),
\end{equation}
where $\eta(v_{\rm min})\equiv \int_{v_{\rm min}}d^3\mathbf{v}_\chi f(\mathbf{v}_\chi)/v_\chi$, and $v_{\rm min}(q,\omega)=\omega/q+q/(2m_\chi)$. As mentioned in the main text, we use the hydrogenic ELF in Eq.~(\ref{eq:ELFfinal}) to describe single-electron ionization from DM scattering with dopants. 
In Fig.~\ref{fig:energyspectrum}, we show the differential rate for DM scattering in a Si:P target with a light dark photon mediator for several DM masses. For all cases, the spectra are peaked at the ionization threshold $E_I=45$ meV and rapidly decrease for larger $\omega$. This feature results from the combination of the kinematics of the light mediator scattering and the ELF. Therefore, the dominant contribution to the total rate comes from the low energy transfer. There is also a sharp cut-off of the spectrum at $\omega_{\rm max}=\frac{1}{2}m_\chi (v_{\rm esc}+v_E)^2$, which is the maximum energy transfer allowed for a given DM mass.

\subsection{Sub-ionization energy depositions and phonon signals}

When the energy transfer to the dopant atom falls below the ionization threshold, dopant electrons are excited into higher-energy bound states that relax back to the ground state by emitting acoustic phonons, which can potentially be measured with calorimetry. To evaluate the DM discovery potential of these sub-ionizing signals, we compute here the contribution of bound-to-bound transitions to DM scattering and absorption rates. 

\begin{figure*}[t!]
\includegraphics[width=0.45\textwidth]{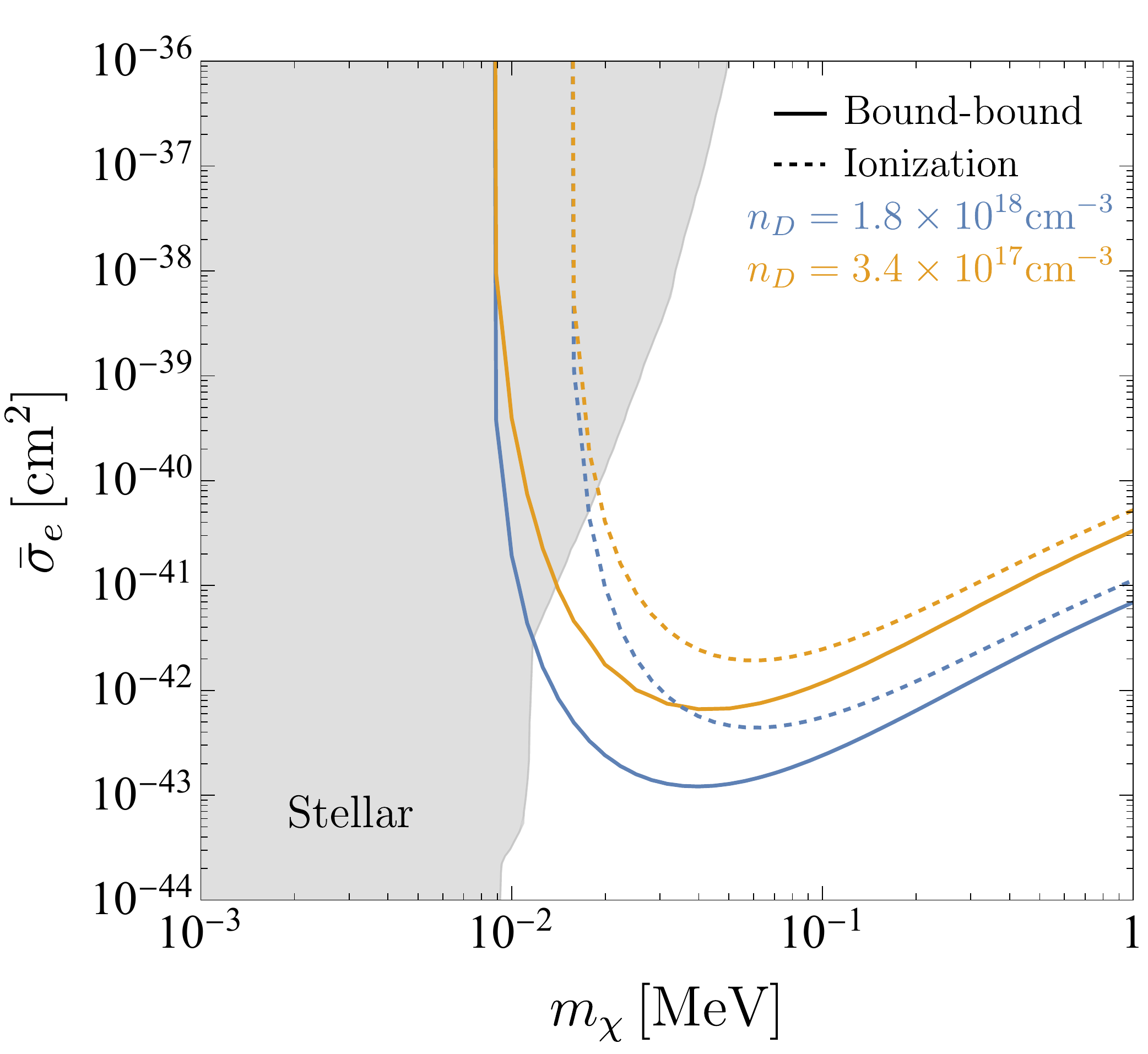}\,\includegraphics[width=0.45\textwidth]{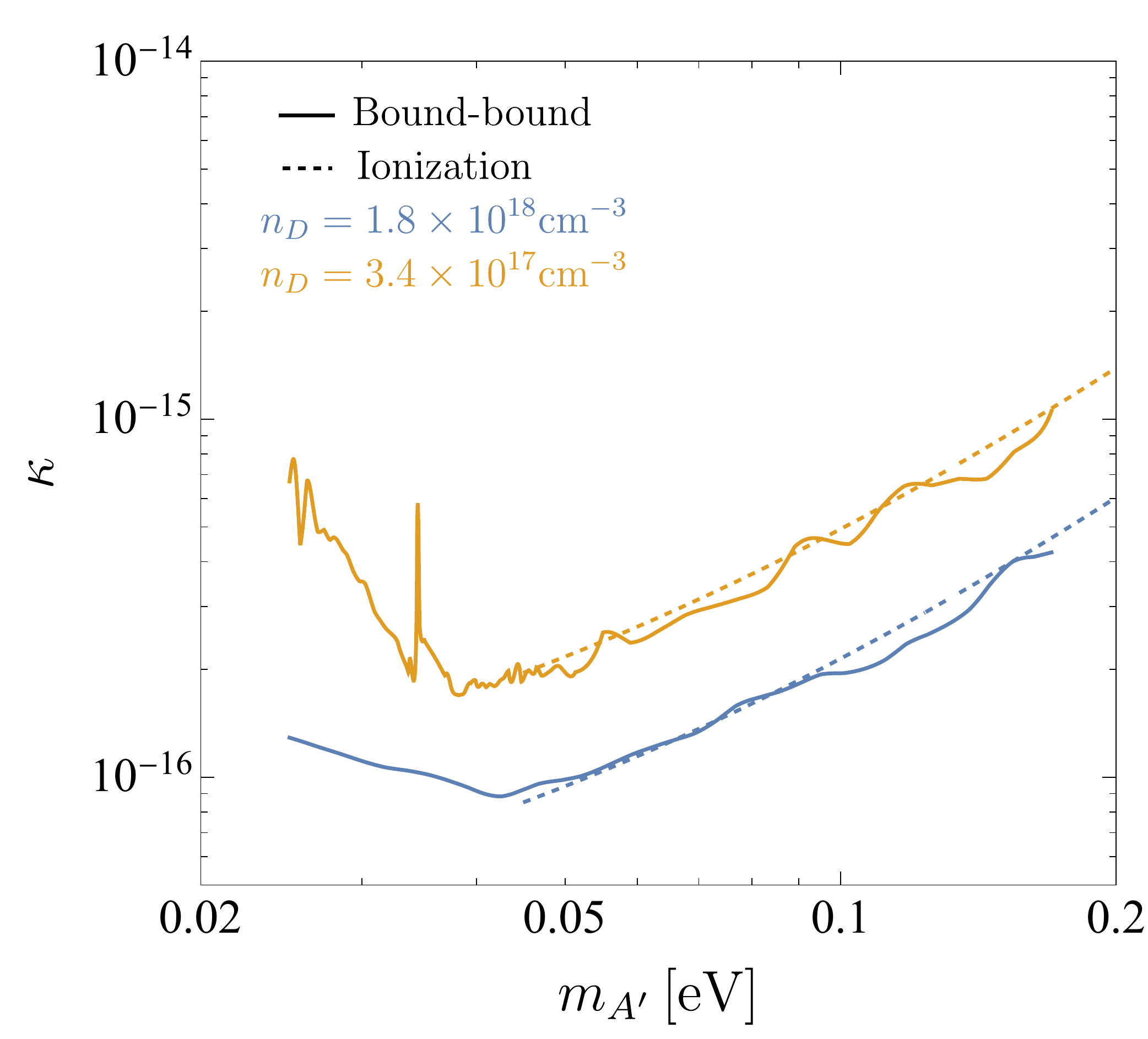}

\caption{\textit{Solid lines:} projected $90\%$ C.L. reach (pure phonon signal from bound-to-bound transitions) of DM-electron scattering with a light dark photon mediator (\textbf{left}) and dark photon DM absorption (\textbf{right}) in Si:P assuming 1 kg-yr exposure with zero background. We present results for two doping densities: $n_D=1.8\times 10^{18}\textrm{cm}^{-3}$ (blue) and $n_D=3.4\times 10^{17}\textrm{cm}^{-3}$ (orange). Bound-to-bound transitions occur for energy depositions  $\omega<E_I=45$ meV, and we have additionally imposed a hard cutoff on the minimum energy transfer of 25 meV due to lack of reliable optical data to compute the ELF for energies below this value. \textit{Dashed lines:} analogous reach for a single ionization signal, \textit{i.e.}, for bound to free electron transitions that require energy depositions above $E_I$. Note that depending on the detection strategy, ionization events can be potentially be detected either by measuring the \textit{phonons} that are released as the ionized electron relaxes to the bottom of the conduction band, or by directly collecting the ionized \textit{electron}, as in standard imaging detectors such as CCDs and BIBs. 
\textit{Gray-shaded region (left panel)} constraint from stellar cooling.
\label{fig:phonon}
}
\end{figure*}

DM interaction rates below the ionization threshold $\omega< E_I$ are not captured by the Hydrogenic ELF Eq.~(\ref{eq:ELFfinal}), which only includes contributions from ionization into free Bloch states. Given the uncertainty in the theoretical description of dopant bound states~\cite{shklovskii2013electronic}, we avoid performing a first-principles computation of the bound-to-bound transitions, and instead simply compute their contribution to the ELF by taking it directly from data of optical absorption for photon energies below the ionization threshold. This amounts to approximating $\mathcal W(\mathbf{q},\omega)\approx\mathcal W (0,\omega)$. Since bound-to-bound transitions are expected for momentum transfers that in Si:P  are at most of order $q\lesssim E_I/v_{\chi}\approx 50$~eV for DM scattering, and $q\lesssim 50$~meV for DM absorption, this approximation is appropriate, as in both cases the momentum transfer lies below the characteristic scale of the dopant atoms which determines the momentum-dependency of the ELF, $qa_*<1$. Using the optical ELF, we obtain rates for DM-induced bound-to-bound transitions for both DM scattering via a light mediator and DM absorption. Assuming that each bound state de-excitation can be measured by collecting the emitted phonons, we show the DM reach, for both scattering and absorption in Si:P in Fig.~\ref{fig:phonon}. We also compare it with the ionization signals discussed in the main text. Thanks to the lower threshold and large bound-to-bound ELF, phonon signals from bound-state de-excitation potentially have a larger sensitivity than ionization signals (by factors of a few), and can probe lighter DM masses. The comparison of the phonon de-excitation and ionization projected reach, however, must be taken lightly,  as we will see below that the technological capabilities and backgrounds of existing single-charge and calorimetric detectors differ substantially. Note that the reach to the phonon signals presented in Fig.~\ref{fig:phonon} is \textit{inclusive}, in the sense that the de-excitation of the bound states likely results in multiple phonons. Here we do not specify the corresponding phonon multiplicity nor spectrum. 

While searching for a small phonon signal may be possible in the future, no detector currently provides sensitivity to single phonons with energies of order $\mathcal O(10)$~meV. In the future, however, two types of sensors may lead to $\mathcal O(10)$~meV single-phonon detection~\cite{Essig:2022dfa}: transition edge sensors (TES) and microwave kinetic inductance detectors (MKID). For these detectors to work, the emitted phonons need to be athermal in the target material, \textit{i.e.}, they must not down-convert into multiple lower energy phonons and must instead travel ballistically within the target, possibly reflecting on its surfaces multiple times before being collected. To our knowledge, no measurement of the phonon lifetime on a doped semiconductor target (with doping densities below the Mott transition) has been made. It is possible, however, that phonons are indeed athermal in doped targets. A doped target has always an irreducible gap, which is the energy required to excite the ground state into the lowest-lying excited state. In Si:P, for instance, the lowest-lying states are the approximately-degenerate 1s states, with the ground state having a ionization energy of order $45$~meV, and the next approximately degenerate state an ionization energy of order $34$~meV~\cite{shklovskii2013electronic}. Thus, the two lowest-lying states have an energy gap  $\approx 10$~meV. As a consequence, in this material we would expect that a phonon emitted from bound-state de-excitations could be reabsorbed in the target if its energy is $\gtrsim 10$~meV, but would travel ballistically for energies below this threshold. The existence of such gaps has been experimentally observed in photoabsorption data. In~\cite{thomas1981optical,PhysRevB.52.16486} it is shown that at low doping densities $n_D \lesssim 10^{16}/\textrm{cm}^3$ and small temperatures $\lesssim 10$~K, doped semiconductors are transparent to photons with energies below $\approx 30$~meV. This stems from the fact that above these energies photons can excite the ground state into the 2p bound state (photonic excitations between the approximately degenerate 1s states are unlikely due to momentum mismatch). At larger doping densities and temperatures the absorption lines broaden significantly due to dispersion of the dopant's energy levels, and the energy gap disappears. While this suggests that lightly-doped semiconductors at cryogenic temperatures may allow for athermal phonons, more experimental and theoretical efforts are required to determine the phonon lifetime on doped targets. 

Regarding backgrounds, calorimetric detectors currently show a large number of unknown events at low energies, which would need to be strongly mitigated for phonon signals to be a viable sub-MeV DM search channel.  For instance, the SuperCDMS CPD experiment ~\cite{SuperCDMS:2020aus} shows a DC rate of approximately $10^4$ events per g-day at low energies, which is a factor of $100$ larger than single-charge detectors discussed previously. While SuperCDMS CPD works in a different energy range than the ones relevant for our proposal (its energy thresholds are of order $10$~eV), backgrounds in different types of calorimetric detectors, including CPD, further increase towards lower energies~\cite{SuperCDMS:2020aus,Fuss:2022fxe}. In~\cite{Anthony-Petersen:2022ujw} it was shown that at least part of the events at TES-based calorimetric detectors are related to stress-induced energy release from auxiliary materials around the detector, but a common background component to detectors subject to large and small levels of stress remains unexplained. Other phonon background in pure semiconductors, which are at present subleading but could be relevant in the future, have been calculated in~\cite{Berghaus:2021wrp}. 

\bibliographystyle{utphys}
\bibliography{Doped_semiconductor}
 
\end{document}